\RequirePackage{ifpdf}
\ifpdf 
\documentclass[pdftex]{sigma}
\else
\documentclass{sigma}
\fi

\usepackage{mathrsfs}
\newtheorem{thm}{Theorem}[section]

\theoremstyle{definition}

\theoremstyle{remark}

\begin{document}

\allowdisplaybreaks

\renewcommand{\thefootnote}{$\star$}

\renewcommand{\PaperNumber}{049}

\FirstPageHeading

\ShortArticleName{Symmetries in Connection Preserving Deformations}
\ArticleName{Symmetries in Connection Preserving Deformations\footnote{This
paper is a contribution to the Proceedings of the Conference ``Symmetries and Integrability of Dif\/ference Equations (SIDE-9)'' (June 14--18, 2010, Varna, Bulgaria). The full collection is available at \href{http://www.emis.de/journals/SIGMA/SIDE-9.html}{http://www.emis.de/journals/SIGMA/SIDE-9.html}}}

\Author{Christopher M.~ORMEROD}

\AuthorNameForHeading{C.M.~Ormerod}

\Address{La Trobe University, Department of Mathematics and Statistics, Bundoora VIC 3086, Australia}
\Email{\href{mailto:C.Ormerod@latrobe.edu.au}{C.Ormerod@latrobe.edu.au}}

\ArticleDates{Received January 31, 2011, in f\/inal form May 18, 2011;  Published online May 24, 2011}

\Abstract{We wish to show that the root lattice of B\"acklund transformations of the $q$-analogue of the third and fourth Painlev\'e equations, which is of type $(A_2+ A_1)^{(1)}$, may be expressed as a quotient of the lattice of connection preserving deformations. Furthermore, we will show various directions in the lattice of connection preserving deformations present equivalent evolution equations under suitable transformations. These transformations correspond to the Dynkin diagram automorphisms.}


\Classification{34M55; 39A13}

\renewcommand{\thefootnote}{\arabic{footnote}}
\setcounter{footnote}{0}

\section{Introduction and outline}

Discrete Painlev\'e equations are non-autonomous second order dif\/ference equations admitting the Painlev\'e equations as continuum limits \cite{Gramani:DiscretePs}. These equations arise as contiguity relations for the Painlev\'e equations \cite{Jimbo:Monodromy2}, from discrete systems arising in quantum gravity \cite{Kitaev:quantumgravity}, reductions of discretizations of classically integrable soliton equations \cite{Hay} and various recurrence relations for orthogonal polynomials~\cite{OrmerodForresterWitte}. There are many ways in which they may be considered integrable such as the singularity conf\/inement property \cite{Gramani:DPainleveProperty}, solvability via associated linear problems \cite{Gramani:Isomonodromic} and algebraic entropy~\cite{Viallet}.

A fundamental result of Okamoto is that the group of B\"acklund transformations of the Painlev\'e equations are of af\/f\/ine Weyl type \cite{Okamoto:Studies:I,Okamoto:Studies:II, Okamoto:Studies:III, Okamoto:Studies:IV}. The work of the Kobe group has remarkably shown the discrete Painlev\'e equations admit similar representations of af\/f\/ine Weyl groups~\cite{Noumi:AffinedPs}. This understanding was enhanced by the pioneering result of Sakai~\cite{Sakai:Rational}, who extended the work of Okamoto on the associated surface of initial conditions to the discrete Painlev\'e equations~\cite{Sakai:Rational}. A valuable insight of this work is that the B\"acklund transformations and the discrete Painlev\'e equations should be considered to be elements of the same group.

It has been well established that the Painlev\'e equations admit Lax representations. The underlying theory behind these Lax representations is the theory of isomonodromy~\cite{Jimbo:Monodromy1, Jimbo:Monodromy2, Jimbo:Monodromy3}. The f\/irst evidence that the $q$-dif\/ference Painlev\'e equations admitted Lax representations can be found in the work of Papageorgiou et al.~\cite{Gramani:Isomonodromic} where it was shown that a $q$-discrete version of the third Painlev\'e equation arises as the compatibility condition of two systems of $q$-dif\/ference equations, written as two $n\times n$ matrix equations of the form
\begin{subequations}
\begin{gather}
\label{Yqx} Y(qx)  = A(x)Y(x),\\
\label{Yqt} \tilde{Y}(x)  = R(x)Y(x),
\end{gather}
\end{subequations}
where $A(x)$ is some $n \times n$ rational matrix in $x$ and $q \in \mathbb{C}$ is a f\/ixed constant such that $|q| \neq 1$ and the evolution denoted by tildes coincides with the evolution of the discrete Painlev\'e equation. If~\eqref{Yqx} and~\eqref{Yqt} form a Lax representation of a Painlev\'e equation, then the compatibility condition determines the evolution of that discrete Painlev\'e equation \cite{Gramani:Isomonodromic}.

The $q$-dif\/ference analogue of the concept of an isomonodromic deformation was proposed later on by Jimbo and Sakai \cite{Sakai:qP6}. The system admits two symbolic solutions,
\begin{gather*}
Y_0(x)  = \hat{Y}_0(x) D_0(x), \\
Y_{\infty}(x)  = \hat{Y}_\infty(x) D_{\infty}(x),
\end{gather*}
where $\hat{Y}_0$ and $\hat{Y}_{\infty}$ are series expansions around $0$ and $\infty$ respectively and $D_0$ and $D_\infty$ are composed of $q$-exponential functions \cite{Sauloy}. Since $Y_0$ and $Y_\infty$ are both fundamental solutions, they should be expressible in terms of each other, which gives rise to the connection matrix,~$C(x)$, specif\/ied by
\begin{gather}\label{connection}
Y_0(x) = Y_\infty(x) C(x).
\end{gather}
For regular systems of linear $q$-dif\/ference equations, such as the linear problem in~\cite{Sakai:qP6}, the solutions converge under more or less general conditions specif\/ied in~\cite{Birkhoffallied}. However, in the irregular case $Y_0(x)$ and $Y_{\infty}(x)$ do not necessarily def\/ine holomorphic functions, hence, we take this def\/inition is taken to be symbolic here. For irregular systems of $q$-dif\/ference equations we have that there exists at least one convergent solution, however, in general, to describe the solutions one would be required to incorporate a $q$-analogue of the Stokes phenomena~\cite{SauloyStokes}.

In this study, we take a connection preserving deformation to be characterized by a transformation of the form~\eqref{Yqt}. In fact, the original Lax representation proposed in \cite{Gramani:Isomonodromic} also def\/ines a~connection preserving deformation in the sense of Jimbo and Sakai \cite{Sakai:qP6}.

In previous studies, we considered a lattice of connection preserving deformations, which is a lattice of shifts of the characteristic data which def\/ines the connection matrix \cite{Ormerodlattice, OrmerodqPV}. However, we struggled to incorporate how the various Dynkin diagram automorphisms \cite{Noumi:AffinedPs} manifested themselves in the theory of the connection preserving deformations. We f\/ind that many connection preserving deformations are copies of the same evolution equations. This induces a natural automorphism on the level of the evolution equations. These automorphisms correspond to Dynkin diagram automorphisms.

We demonstrate the role of the Dynkin diagram automorphisms in a study of the associated linear problem for $q$-$P_{\rm III}$ and $q$-$P_{\rm VI}$, which possesses a group of B\"acklund transformations of type $W((A_2+A_1)^{(1)})$, which is the case of $P(A_5^{(1)})$ in Sakai's notation. This is an interesting case as the symmetries have been nicely studied before~\cite{NoumiqP4} and that the two equations possess the same surface of initial conditions~\cite{Sakai:Rational} and the same associated linear problem where the two systems are dif\/ferent directions on the lattice of connection preserving deformations~\cite{Ormerodlattice}. This lattice of connection preserving deformations admits a natural group of automorphisms of order six, corresponding naturally to the sub-group of B\"acklund transformations generated by the Dynkin diagram automorphisms. This advances our understanding of the relation between the associated linear problems for $q$-dif\/ference equations and their symmetries.

\section{Birkhof\/f theory for irregular dif\/ference equations}

Before specifying the form of the solutions, let us f\/ix some notation required to specify the solutions. We specify some of the building blocks used in the Galois theory approach to the study of systems of linear $q$-dif\/ference equations \cite{Sauloy, vanderPut}. Let us consider the $q$-Pochhammer symbol, given by
\[
(a;q)_k = \left\{
\begin{array}{ll}
  \prod\limits_{n=0}^{k-1} (1-aq^n) & {\rm if} \  0 < k < \infty ,\vspace{1mm}\\
1 & {\rm if} \ k = 0,\vspace{1mm}\\
  \prod\limits_{n=0}^{\infty} (1-aq^n) & {\rm if} \ k = \infty,
\end{array}\right.
\]
where we will also use the notation
\[
(a_1,\ldots,a_n;q)_k = \prod_{i = 1}^n (a_i;q)_k.
\]
We specify the Jacobi theta function as
\[
\theta_q(x) = \sum_{n\in\mathbb{Z}} q^{{n \choose 2}} x^n,
\]
which satisf\/ies
\[
x\theta_{q}(qx)=\theta_q(x).
\]
It is also useful to def\/ine the $q$-character as
\[
e_{q,c}(x) = \frac{\theta_{q}(x)\theta_{q}(c)}{\theta_{q}(xc)},
\]
which consequently satisf\/ies the pair of equations
\begin{gather*}
e_{q,c}(qx) = c e_{q,c}(x), \qquad e_{q,qc}(x) = x e_{q,c}(x).
\end{gather*}
Using the above building blocks, it is an elementary task to construct solutions to any f\/irst order linear $q$-dif\/ference equation, hence, we may transform any system of $q$-dif\/ference equations of the form~\eqref{Yqx}, where $A(x)$ is rational, to a system of $q$-dif\/ference equations where $A(x)$ is polynomial. Hence, without loss of generality, we may let
\[
A(x) = A_0 + A_1 x + \dots + A_m x^m.
\]
We isolate the set of points in which $A(x)$ is singular, i.e., where $\det A(x) = 0$, by f\/ixing the determinant as
\begin{gather*}
\det A(x) = \kappa x^L (x-a_1)\cdots (x-a_r).
\end{gather*}
We will call a problem of the form \eqref{Yqx} regular if $A_0$ and $A_m$ are diagonalizable and invertible and irregular if it is not regular. 
The formal expansion for regular problems may be computed under certain non-resonance conditions specif\/ied by the work of Birkhof\/f~\cite{Birkhoffallied}. However, to pass to the irregular theory, we must refer to the work of Adams~\cite{Adams}, which was subsequently rediscovered by Birkhof\/f and Guenther~\cite{BirkhoddAdamsSum}.

\begin{thm}
The linear problem \eqref{Yqx} admits two fundamental series solutions, $Y_{0}(x)$ and $Y_{\infty}(x)$, specified by the expansions
\begin{subequations}
\begin{gather}
\label{y0}Y_{0}(x)  = \left( Y_0 + Y_{1}x + Y_2x^2 + \cdots \right) \mathrm{diag}\left( \dfrac{e_{q,\lambda_i}(x)}{\theta(x)^{l_i}}\right) , \\
\label{yinf}Y_{\infty}(x)  = \left( Y_0' + \frac{Y_{-1}}{x} + \frac{Y_{-2}}{x^2} + \cdots \right) \mathrm{diag} \left( \dfrac{e_{q,\kappa_i}(x)}{\theta(x)^{k_i}}\right),
\end{gather}
\end{subequations}
where $\kappa_i$, $\lambda_i$, $l_i$ and $k_i$ must satisfy the condition that if $l_i = l_j$ $(k_i = k_j)$ then $\lambda_i \neq \lambda_jq^p$ $(\kappa_i \neq \kappa_jq^p)$ for any integer $p \neq 0$.
\end{thm}

In general, we are required to choose $Y_0$ and $Y_0'$ to diagonalize $A_0$ and $A_m$ respectively. More details on the process of f\/inding these solutions may be found in the work of Adams~\cite{Adams}. In this irregular setting the solutions found do not necessarily def\/ine holomorphic functions~\cite{BirkhoffTrjitzinsky}. Once these solutions are def\/ined, we may specify the connection matrix via~\eqref{connection}. However, we are required to take \eqref{connection} to be a symbolic def\/inition at this point.

To characterize the set of deformations considered, we introduce the characteristic data \cite{Ormerodlattice, OrmerodqPV}. The characteristic data consists of the variables def\/ining the asymptotic behavior of the solutions at $x= 0$ and $x= \infty$, def\/ined by~\eqref{y0} and~\eqref{yinf}, and the roots and poles of $A(x)$. We denote this data by
\[
M  =\left\{ \begin{array}{c c } \kappa_1, \ldots, \kappa_n & a_1, \ldots, a_r \\ \lambda_1, \ldots, \lambda_n & \end{array} \right\}.
\]
However, while this study extends previous works \cite{Ormerodlattice, OrmerodqPV}, a task remains to fully describe the set of solutions by incorporating a $q$-analogue of the Stokes phenomenon \cite{SauloyStokes}. That is, in order to fully mirror the theory of monodromy, we are also required to include data that encodes the Stokes phenomenon for systems of linear $q$-dif\/ference equations \cite{SauloyStokes}.

We now take a deformation of \eqref{Yqx} to be def\/ined by \eqref{Yqt}, where it is easy to see that $\tilde{Y}(x)$ must satisfy
\begin{gather}\label{newev}
\tilde{Y}(qx) = \left[R(qx)A(x)R(x)^{-1} \right] \tilde{Y}(x) = \tilde{A}(x) \tilde{Y}(x),
\end{gather}
which def\/ines $\tilde{A}(x)$. We take this new linear problem to be associated with a new connection matrix, $\tilde{C}(x)$, and a new set of characteristic data, denoted $\tilde{M}$. If the fundamental solutions satisfy the conditions
\begin{subequations}
\begin{gather}
\label{y0change}\tilde{Y}_{0}(x)  = R(x) Y_0(x), \\
\label{yinfchange}\tilde{Y}_{\infty}(x)  = R(x) Y_{\infty}(x),
\end{gather}
\end{subequations}
then it is clear from \eqref{connection} that \eqref{y0change} and~\eqref{yinfchange} implies that $\tilde{C}(x) = C(x)$. In the regular case, a~specif\/ic case of the converse implication was presented in the work Jimbo and Sakai's work~\cite{Sakai:qP6}, however, for the case of irregular systems of $q$-dif\/ference equations, where the fundamental solutions are not necessarily holomorphic functions, the analogous converse implication is not so clear.

If we do take $R(x)$ to be rational, as reported in a previous work~\cite{Ormerodlattice}, we can say that $\tilde{M}$ has an altered set of characteristic constants. On the level of the series solutions of Adams~\cite{Adams}, Birkhof\/f and Guenther~\cite{BirkhoddAdamsSum}, a transformation of the form~\eqref{Yqt} may
\begin{itemize}\itemsep=0pt
\item change the asymptotic behavior of the fundamental solutions at $x = \infty$ by letting $\kappa_i \to q^{n}\kappa_i$;
\item change the asymptotic behavior of the fundamental solutions at $x = 0$ by letting $\lambda_i \to q^{n}\lambda_i$;
\item change the position of a root of the determinant by letting $a_i \to q^na_i$.
\end{itemize}
We cannot deform these variables arbitrarily, as there is one constraint we must satisfy; if we consider the determinant of the left and right hand side of \eqref{Yqx} for the solution $Y= Y_0$ at $x= 0$, it is easily shown that
\begin{gather}\label{constraint}
\prod \kappa_i \prod (-a_i) = \prod \lambda_i.
\end{gather}
This must also hold true for $\tilde{M}$. This is a constraint on how we may deform the characteristic data.

Conversely, let us suppose that $\lambda_i$ and $\kappa_i$ are changed by a multiplication by some power of~$q$, then $\tilde{D}_0 D_0^{-1}$ and $\tilde{D}_\infty D_\infty^{-1}$ are both rational functions, meaning that the expansions
\begin{gather}\label{expansions}
R(x)  =   \tilde{Y}_0(x) Y_0(x)^{-1},\qquad
R(x)  =    \tilde{Y}_\infty(x) Y_\infty(x)^{-1},
\end{gather}
present two dif\/ferent expansions around $x= 0$ and $x= \infty$ respectively. Furthermore, taking a~determinant of~\eqref{newev} as it def\/ines $\tilde{A}(x)$ specif\/ies that $R(x)$ satisf\/ies the equation
\[
\dfrac{\det R(qx)}{\det R(x)} = \dfrac{\det \tilde{A}(x)}{\det A(x)}.
\]
An alternative characterization of \eqref{newev} is given by the way in which we have two ways to calculate~$\tilde{Y}(qx)$:
\begin{gather*}
\tilde{Y}(qx)  =  \tilde{A}(x)R(x)Y(x),\qquad
\tilde{Y}(qx)  = R(qx)A(x)Y(x),
\end{gather*}
leading to the compatibility condition
\[
\tilde{A}(x)R(x) = R(qx)A(x).
\]
This type of condition appears throughout the integrable literature \cite{Gramani:Isomonodromic, Sakai:qP6}. Given a~deformation of the characteristic constants,~$\tilde{M}$, the above constitutes enough information to determine $R(x)$ and also determine the transformation $R(x)$ induces~\cite{Ormerodlattice}.

At this point, we wish to outline the idea of the symmetries of the associated linear problem as one f\/inds a host of deformations of the characteristic constants, which we call the lattice of connection preserving deformations, presented in a~previous work~\cite{Ormerodlattice}. This is an idea distinct from the work of Jimbo and Sakai~\cite{Sakai:qP6} in that there is no one canonical deformation, but rather a family of them. If one considers the set of all possible moves, one may endow this with a~lattice structure of dimension $r+ 2n -1$. In addition to the above translations one expects a~certain number of symmetries to be natural in this setting. One of the most natural groups that should be allowed to act on the set of characteristic data should be the group of permutations on the~$r$ roots \cite{Ormerodlattice}.

We wish to examine an additional structure that may be obtained from the lattice. We consider two systems to be equivalent if there exists a transformation that maps the evolution of one system to the other and vice versa. A classical example would be $P_{34}$ and $P_{\rm II}$, which are related via a Miura transformation. We will f\/ind that the transformations induced by many of the directions in the lattice of connection preserving deformations are equivalent, splitting the group into classes of equivalent systems, related to each other by a transformation. Conceptually, one may think of these transformations that permute the directions of the lattice as rotations on the lattice, and that the fact that the lattice admits these rotations as being a property of the lattice.

\section{The associated linear problem}

We will begin by specifying the properties of the linear system:
\begin{itemize}\itemsep=0pt
\item The matrix $A(x)$ is chosen so that
\begin{gather*}
\det A(x) = \kappa_1\kappa_2x (x-a_1)(x-a_2).
\end{gather*}

\item The solution at $x = 0$ is given by
\begin{gather}\label{III0form}
Y_{0}(x) = \left( Y_0 + Y_{1}x + Y_2x^2 + \cdots \right)\begin{pmatrix} e_{q,\lambda_1}(x) & 0 \\ 0 &  e_{q,\lambda_2}(x)/\theta_q(x)\end{pmatrix}.
\end{gather}

\item The solution at $x = \infty$ is given by
\begin{gather}\label{IIIinfform}
Y_{\infty}(x) = \left( I + \dfrac{Y_{-1}}{x} + \dfrac{Y_{-2}}{x^2} + \cdots \right)\begin{pmatrix} e_{q,\kappa_1}(x)/\theta_q(x)^2 & 0 \\ 0 & e_{q,\kappa_2}(x)/\theta_q(x)\end{pmatrix}.
\end{gather}
\end{itemize}

The determinant and the various other asymptotics specify that
\[
A(x) = A_0 + A_1x + A_2x^2.
\]
This would ordinarily leave us to def\/ine twelve variables, however, the normalization of the solution at $x = \infty$, the determinantal constraint and the solutions at $0$, in addition to the relation
\[
\kappa_1\kappa_2a_1a_2 = \lambda_1 \lambda_2
\]
leave three variables to be chosen. If one choose the variables arbitrarily, the evolution equations are not guaranteed to be nice. So the question remains, how does one choose these variables so that the resulting evolution equations appear in some sort of canonical manner? It is here that the continuous case lends incredible insight: if one refers to the work of Jimbo, Miwa and Ueno~\cite{Jimbo:Monodromy1, Jimbo:Monodromy2, Jimbo:Monodromy3}, it is apparent that the relevant variables, almost invariably are chosen so that one variable is the zero of the upper right entry of the matrix, one variable encapsulates the gauge freedom and one more variable is associated with the evaluation of the diagonal elements at the root of the upper right entry. That is to say that if we def\/ine a notation for the individual entries of $A(x)$ as $A(x) = (a_{i,j}(x))$, then $a_{1,2}(y) = 0$. Furthermore, the gauge freedom is encapsulated by letting
\[
a_{1,2}(x) \propto w(x-y).
\]
There remains one more variable to choose, and this is done so that
\[
A(y) = \begin{pmatrix} z_1 & 0 \\ * & z_2 \end{pmatrix},
\]
where there is a determinantal constraint linking $z_1$ to $z_2$. In the continuous case, it is typical that this constraint is that $A(x)$ is traceless at $x=y$, giving $z_1 = -z_2$. In our case, we will require that $\det(A(y)) = z_1 z_2$, which specif\/ies one degree of freedom, chosen to be represented by a variable, $z$. Remarkably, this choice seems to be canonical in that the resulting evolution equations, after one performs the appropriate connection preserving deformations, appear to be in a symplectic form.

This completely specif\/ies a matrix parameterization of the form
\begin{gather*}
A(x) = \begin{pmatrix} \kappa_1((x-y)(x-\alpha) + z_1 & \kappa_2 w (x-y) \vspace{1mm}\\
\dfrac{\kappa_1}{w} (\gamma x + \delta) & \kappa_2(x-y + z_2)
\end{pmatrix},
\end{gather*}
where $\alpha$, $\gamma$ and $\delta$ are functions of $y$ and $z$ specif\/ied by the constraints. As it is quite easily derivable from the above constraints, and variants have appeared before \cite{Murata2009, Ormerodlattice, OrmerodqPV}, we simply list the parameters as
\begin{gather*}
\alpha  =\frac{-z_1 \kappa _1+\left(y-z_2\right) \kappa _2+\lambda _1}{y \kappa _1},\qquad
\gamma  =-2 y-\alpha +a_1+a_2+z_2,\\
\delta  =\frac{\left(y \alpha +z_1\right) \left(y-z_2\right)}{y},\qquad
z_1 =\frac{y \left(y-a_1\right)}{z},\qquad
z_2 =z \left(y-a_2\right).
\end{gather*}
Now that we have a parameterization for $A(x)$, to f\/ind the relevant connection preserving deformations using \eqref{expansions}, we require knowledge of the fundamental solutions. To do this, we simply substitute~\eqref{III0form} and~\eqref{IIIinfform} into~\eqref{Yqx} and solve for~$Y_i$. It is suf\/f\/icient to compute the f\/irst few terms for computation of the connection preserving deformations specif\/ied later on. We will start with the expansion around~$x= \infty$. Let us f\/irst specify notation as $Y_k = \big(y_{i,j}^{(k)}\big)$, then if we consider $0 =Y_{\infty}(qx) - A(x)Y_{\infty}$, at order~$x^{-3}$ we have
\[
0 = \begin{pmatrix}
 \dfrac{x \kappa _1 \big(q (y+\alpha )-(q-1) y_{1,1}^{(-1)}\big)}{q} & x \big(-w \kappa _2-\kappa _1 y_{1,2}^{(-1)}\big) \vspace{2mm}\\
 \dfrac{x \kappa _1 \big(w y_{2,1}^{(-1)}-q \gamma \big)}{q w} & \dfrac{w \kappa _2 \big(q y-q z_2-(q-1) y_{2,2}^{(-1)}\big)-q \gamma  \kappa _1 y_{1,2}^{(-1)}}{q w}
 \end{pmatrix}
\]
specifying $Y_{-1}$ completely. We could go on to calculate $Y_{-2}$ and $Y_{-3}$, however, in the interest of keeping this account elegant, we shall stop. However, it suf\/f\/ices to say that the series solution may be found to any order.

To specify the solution around $x=0$, we f\/irst need to def\/ine $Y_0$. Of course, there is a non-uniqueness in how we def\/ine $Y_0$, as given any matrix that diagonalizes $A_0$, multiplication on the right by any invertible diagonal matrix results in another matrix that diagonalizes $A_0$. We choose a relatively simple matrix to simplify some of our calculations, namely we choose $Y_0$ to be
\[
Y_0 = \begin{pmatrix} w y & w y \kappa _2 \\
 -z y+y+z a_2 & \left(-z y+y+z a_2\right) \kappa _2+\lambda _1
\end{pmatrix}.
\]
Although it is quite simple to compute $Y_1$, $Y_2$ and so on, the computations become increasingly verbose, hence, we will only list the above terms.

\section{The symmetries of the associated linear problem\\ and the B\"acklund transformations}

Now that everything has been def\/ined, we may study the lattice of connection preserving deformations. We have six variables in total that we may deform and one constraint, specif\/ied by~\eqref{constraint}. This gives us a f\/ive basis elements to f\/ind. To reduce some of the workload, let us consider the most basic of transformation: notice that if we multiply the fundamental solutions by $R(x) = xI$, where $I$ is simply the identity matrix, we may absorb this into the factors, $D_0$~and~$D_{\infty}$, by letting the $\kappa_i \to \tilde{\kappa}_i = q \kappa_i$ and $\lambda_i \to \tilde{\lambda}_i = q \lambda_i$. Since this does not change the relative gauge freedom, this does not change the def\/inition of $y$, hence, $\tilde{y} = y$, $\tilde{z} = z$ and $\tilde{w} = \tilde{w}$. We f\/ind it convenient to introduce the notation
\[
T_{\kappa_1,\kappa_2,\lambda_1,\lambda_2} : \ \left\{ \begin{array}{c c c} \kappa_1 & \kappa_2 & a_1 \\ \lambda_1 & \lambda_2 & a_2 \end{array} : w,y,z \right\} \to \left\{ \begin{array}{c c c} q\kappa_1 & q\kappa_2 & a_1 \\ q\lambda_1 & q\lambda_2 & a_2 \end{array} : \tilde{w},\tilde{y},\tilde{z} \right\},
\]
where  $\tilde{w} = w$, $\tilde{y}= y$ and $\tilde{z}= z$ in accordance with the transformation. We need to choose four separate generators.

To reduce workload once more, it is clear, in accordance with the parameterization, there is a symmetry we may exploit. As mentioned above, we expect to be able to permute the roots in a natural manner, if we appropriately def\/ine $\tilde{y}$, $\tilde{z}$ and $\tilde{w}$. We note that $\tilde{A}(x) = A(x)$ if we let
\begin{gather*}
S_{a_1,a_2} :  \ \left\{ \begin{array}{c c c} \kappa_1 & \kappa_2 & a_1 \\ \lambda_1 & \lambda_2 & a_2 \end{array} : w,y,z \right\} \to \left\{ \begin{array}{c c c} \kappa_1 & \kappa_2 & a_2 \\ \lambda_1 & \lambda_2 & a_1 \end{array} : w,y,z\dfrac{y-a_2}{y-a_1} \right\}.
\end{gather*}
We choose the following additional elements to represent the f\/ive dimensional lattice:
\begin{gather*}
T_{a_1,\lambda_1}  : \ \left\{ \begin{array}{c c c} \kappa_1 & \kappa_2 & a_1 \\ \lambda_1 & \lambda_2 & a_2 \end{array} : w,y,z \right\} \to \left\{ \begin{array}{c c c} \kappa_1 & \kappa_2 & qa_1 \\ q\lambda_1 & \lambda_2 & a_2 \end{array} : \tilde{w},\tilde{y},\tilde{z} \right\},\\
T_{a_2,\lambda_1}  : \ \left\{ \begin{array}{c c c} \kappa_1 & \kappa_2 & a_1 \\ \lambda_1 & \lambda_2 & a_2 \end{array} : w,y,z \right\} \to \left\{ \begin{array}{c c c} \kappa_1 & \kappa_2 & a_1 \\ q\lambda_1 & \lambda_2 & qa_2 \end{array} : \tilde{w},\tilde{y},\tilde{z} \right\},\\
T_{\kappa_1,\lambda_1}  : \  \left\{ \begin{array}{c c c} \kappa_1 & \kappa_2 & a_1 \\ \lambda_1 & \lambda_2 & a_2 \end{array} : w,y,z \right\} \to \left\{ \begin{array}{c c c} q\kappa_1 & \kappa_2 & a_1 \\ q\lambda_1 & \lambda_2 & a_2 \end{array} : \tilde{w},\tilde{y},\tilde{z} \right\},\\
T_{\kappa_2,\lambda_2}  : \ \left\{ \begin{array}{c c c} \kappa_1 & \kappa_2 & a_1 \\ \lambda_1 & \lambda_2 & a_2 \end{array} : w,y,z \right\} \to \left\{ \begin{array}{c c c} q\kappa_1 & \kappa_2 & a_1 \\ \lambda_1 & q\lambda_2 & a_2 \end{array} : \tilde{w},\tilde{y},\tilde{z} \right\},
\end{gather*}
where $\tilde{w}$, $\tilde{y}$ and $\tilde{z}$ is to be determined in each case. These four elements and $T_{\kappa_1,\kappa_2,\lambda_1,\lambda_2}$ form a~basis for the lattice of connection preserving deformations.

\begin{thm}
The transformations $T_{a_1,\lambda_1}$, $T_{\kappa_1,\lambda_1}$ and $T_{\kappa_1,\lambda_2}$ are induced by transformations of the linear problem
\begin{gather*}
\tilde{Y}(x)  = R_{a_1,\lambda_1}(x) Y(x),\qquad
\tilde{Y}(x)  = R_{\kappa_1,\lambda_1}(x) Y(x),\qquad
\tilde{Y}(x)  = R_{\kappa_1,\lambda_2}(x) Y(x),
\end{gather*}
respectively, where
\begin{subequations}
\begin{gather}
\label{Ra1l1}
R_{a_1,\lambda_1}  = \begin{pmatrix}
 \dfrac{x+q y-q a_1+\dfrac{q y \left(y-a_1\right)}{z \left(a_2-y\right)}}{x-q a_1} & \dfrac{q w y \left(y-a_1\right)}{z \left(x-q
   a_1\right) \left(y-a_2\right)} \vspace{2mm}\\
 -\dfrac{q \left(y (z-1)-z a_2\right) \left(y (z-1)+a_1-z a_2\right)}{w z \left(x-q a_1\right) \left(y-a_2\right)} & \dfrac{x+q y
   \left(\frac{y-a_1}{y z-z a_2}-1\right)}{x-q a_1}\end{pmatrix},\\
\label{Rk1l1}
R_{\kappa_1,\lambda_1}(x)  = \begin{pmatrix} x+\dfrac{\left(y (z-1)-z a_2\right) \kappa _2}{y \kappa _1} & \dfrac{w \kappa _2}{\kappa _1}
\vspace{2mm}\\ \dfrac{y (z-1)-z a_2}{w y} & 1 \end{pmatrix},\\
\label{Rk1l2}
R_{\kappa_1,\lambda_2}(x)  = \begin{pmatrix}
 x+\dfrac{\left(y (z-1)-z a_2\right) \kappa _2-\lambda _1}{y \kappa _1} & \dfrac{w \kappa _2}{\kappa _1} \vspace{2mm}\\
 \dfrac{\left(y (z-1)-z a_2\right) \kappa _2-\lambda _1}{w y \kappa _2} & 1
\end{pmatrix}.
\end{gather}
\end{subequations}
\end{thm}

\begin{proof}
In considering $R_{a_1,\lambda_1}$, since $\kappa_1$ and $\kappa_2$ are unchanged, we know that
\begin{gather*}
\tilde{Y}_{\infty}(x)  Y_{\infty}(x)^{-1} = I + O\left( \dfrac{1}{x}\right) ,
\end{gather*}
this, coupled with the determinantal relation specify that
\[
\det R_{a_1,\lambda_1} = \frac{x}{x-a_1},
\]
hence, we seek a parameterization of the form
\[
R_{a_1,\lambda_1}(x) = \frac{xI + R_0}{x-qa_1}.
\]
We consider the compatibility in determining $\tilde{Y}(qx)$ to be given by $R(qx)A(x) = \tilde{A}(x)R(x)$. Taking the residue at $x= a_1$ gives us $R_0$ completely in terms of the untransformed variables, namely, we obtain a representation of $R_{a_1,\lambda_1}$ given by \eqref{Ra1l1}.

To compute $R_{\kappa_1,\lambda_1}(x)$, the dif\/ference in asymptotic behavior of $Y_\infty(x)$ with $\tilde{Y}_\infty(x)$ is used as we notice
\begin{gather*}
\tilde{Y}_{\infty}(x)  Y_{\infty}(x)^{-1}  = \tilde{\hat{Y}}_\infty(x) \tilde{D}_{\infty}(x) D_{\infty}(x)^{-1} \hat{Y}_{\infty}^{-1}
 = \tilde{\hat{Y}}_\infty(x) \begin{pmatrix} x & 0 \\ 0 & 1\end{pmatrix} \hat{Y}_{\infty}^{-1}.
\end{gather*}
The leading terms of $\hat{Y}_{\infty}(x)$ and $\tilde{\hat{Y}}_{\infty}(x)$ are both $I$, hence, $R_{\kappa_1,\lambda_1}$ is given by a formal expansion of the form
\[
R_{\kappa_1,\lambda_1}(x) = x\begin{pmatrix} 1 & 0 \\ 0 & 0 \end{pmatrix} + R_1 + O\left( \dfrac{1}{x} \right),
\]
where we use the fact $\det R_{\kappa_1,\lambda_1}(x) = x$ to bound the order of the expansion. Where the previous calculation allowed a simple derivation of the entries of $R_{a_1,\lambda_1}(x)$ via the computation of the residue, there is an added dif\/f\/iculty in computing  $R_{\kappa_1,\lambda_1}$. We are required to look at va\-rious combinations of the entries of the compatibility relation,  $\tilde{A}(x)R_{\kappa_1,\lambda_1}(x) = R_{\kappa_1,\lambda_1}(qx)A(x)$. However, it is not a very dif\/f\/icult calculation. The result is given by~\eqref{Rk1l1}. The derivation of~$R_{\kappa_1,\lambda_2}$ follows the same pattern as the previous case.
\end{proof}

While we have mainly used the compatibility relation to def\/ine the entries of the $R$ matrices, it is also possible to expand the $\tilde{Y}_\infty(x) (Y_\infty(x))^{-1}$ to higher orders to f\/ind $R$. However, however one may determine the entries, it becomes abundantly clear that we have many more relations than we need to def\/ine the entries of $R$ in each case. The remaining relations may be used to express $\tilde{y}$, $\tilde{z}$ and $\tilde{w}$ in terms of $y$, $z$ and $w$.

\begin{thm}
The effect of the transformations, $T_{a_1,\lambda_1}$, $T_{\kappa_1,\lambda_1}$ and $T_{\kappa_1,\lambda_2}$, are specified by the relations
\begin{alignat*}{3}
& T_{a_1,\lambda_1} : \ && \tilde{w}=w \left(1-\tilde{z}\right), \qquad
 \tilde{y}y =\dfrac{\tilde{z} \left(\tilde{z} a_2 \kappa _2+q \lambda _1\right)}{q \left(\tilde{z}-1\right) \kappa _1},\qquad
  \tilde{z}z =\dfrac{q y \left(y-a_1\right) \kappa _1}{\left(y-a_2\right) \kappa _2}, &\\
& T_{\kappa_1,\lambda_1}  : \ && \tilde{w}=\dfrac{w \left(\tilde{z}-1\right) \left(q y \kappa _1-z \kappa _2\right)}{z \kappa _1},\qquad
  \tilde{y}y =\dfrac{z \left(\tilde{z} a_2 \kappa _2+q \lambda _1\right)}{q \left(\tilde{z}-1\right) \kappa _1},\qquad
 \tilde{z} =\dfrac{q y a_1 \kappa _1+q z \lambda _1}{q y^2 \kappa _1-y z \kappa _2},& \\
& T_{\kappa_1,\lambda_1}  : \ \ && \tilde{w}=-w \left(\tilde{z}-1\right),\qquad
  y \tilde{y}=\dfrac{\tilde{z} \left(\tilde{z} a_2 \kappa _2+q \lambda _1\right)}{q \left(\tilde{z}-1\right) \kappa _1},& \\
&&& \tilde{z}z =-\dfrac{q \kappa_1 \left(z a_1 a_2 \kappa _2+\left(a_1-y\right) \lambda _1\right)}{ \kappa _2 \left(a_2 \left(\kappa _2-q y \kappa _1\right)+\lambda _1\right)}. &
\end{alignat*}
\end{thm}

Note that inverses of all these transformations are easily computed in the form presented. We now state that the lattice of connection preserving deformations as
\[
\mathscr{L} = \langle T_{\kappa_1,\kappa_2,\lambda_1,\lambda_2}, T_{a_1,\lambda_1} , T_{a_2,\lambda_1} , T_{\kappa_1,\lambda_1} , T_{\kappa_1,\lambda_2}\rangle \cong \mathbb{Z}^5,
\]
the elements above commute and form a basis for this lattice.

\section[The multiple correspondences to $q$-$P_{\rm III}$ and $q$-$P_{\rm IV}$]{The multiple correspondences to $\boldsymbol{q}$-$\boldsymbol{P_{\rm III}}$ and $\boldsymbol{q}$-$\boldsymbol{P_{\rm IV}}$}

There are two problems we wish to address in this section that are not covered by previous studies \cite{Ormerodlattice}. Firstly, the dimension of the lattice of connection preserving deformations is f\/ive, whereas the root lattice of type $(A_2+A_1)^{(1)}$ is three. We also wish to look at how the various correspondences between the two lattices gives rise to the Dynkin diagram automorphisms.

One expects that the set of translational B\"acklund transformations embed naturally in this lattice of connection preserving deformations. We introduce a similar notation on the variables for the af\/f\/ine Weyl group of type $(A_2+A_1)^{(1)}$, as found in the work of Kajiwara et al.~\cite{KajiwaraqP3I}. We def\/ine one ref\/lection, $s_1$, and one rotation, $\sigma_0$, which generates a group of type $A_2^{(1)}$, given by
\begin{gather*}
s_0  : \ \left\{\begin{array}{c c}  b_0 & b_1 \\ b_2 &  \end{array} f_0,f_1,f_2\right\} \to \left\{\begin{array}{c c}  \frac{1}{b_0} & b_1b_0 \\ b_2b_0 &  \end{array} f_0, f_1\left(\frac{b_0+f_0}{1+b_0f_0} \right), f_2\left( \frac{1+b_0f_0}{b_0+f_0}\right)\right\},\\
\sigma_0  : \ \left\{\begin{array}{c c}  b_1 & b_2 \\ b_0 &  \end{array} f_1, f_2, f_0\right\} \to \left\{\begin{array}{c c}  b_0 & b_1 \\ b_2 &  \end{array} f_0, f_1, f_2\right\},
\end{gather*}
where the other ref\/lections may be def\/ined to be $s_1 = \sigma_0 \circ s_0 \circ \sigma_0^2$ and $s_2 = \sigma_0^2 \circ s_0 \circ \sigma_0$. We also def\/ine some extra generators, $\sigma_1$ and $w_0$, to be
\begin{gather*}
\sigma_1 : \ \left\{\begin{array}{c c}  b_0 & b_1 \\ b_2 &  \end{array} f_0, f_1, f_2\right\} \to \left\{\begin{array}{c c}  b_0 & b_1 \\ b_2 &  \end{array} \frac{1}{f_0}, \frac{1}{f_1}, \frac{1}{f_2}\right\},\\
w_0  : \ \left\{\begin{array}{c c}  b_0 & b_1 \\ b_2 &  \end{array} f_0, f_1, f_2\right\} \to \left\{\begin{array}{c c} b_0 & b_1 \\ b_2 &  \end{array} \frac{b_0b_1(b_2b_0+b_2f_0+f_2f_0)}{f_2(b_0b_1+b_0f_1+f_0f_1)} ,\right. \\
\qquad \ \ \left. \frac{b_1b_2(b_0b_1+b_0f_1+f_0f_1)}{f_0(b_1b_2+b_1f_2+f_1f_2)}, \frac{b_1b_2(b_0b_1+b_0f_1+f_0f_1)}{f_0(b_1b_2+b_1f_2+f_1f_2)}\right\},
\end{gather*}
where we def\/ine one more operator, $w_1 = \sigma_1 \circ w_0 \circ \sigma_1$. It is a general result of~\cite{KajiwaraqP3I} that the generators of $G = \langle s_0, s_1,s_2,w_0,w_1,\sigma_0, \sigma_1 \rangle$ satisfy all the relations of the extended af\/f\/ine Weyl group of type $(A_2+A_1)^{(1)}$.

Following \cite{KajiwaraqP3I}, there are four translational components, $T_0, \ldots, T_3$, where $T_0 \circ T_1 \circ T_2 = I$, which generates a three dimensional lattice. These are
\[
T_0 = \sigma_0 \circ s_2 \circ s_1,\qquad T_1 = \sigma_0 \circ T_1 \circ \sigma_0^2,\qquad T_2 = \sigma_0^2 \circ T_1 \circ \sigma_0, \qquad T_3 = \sigma_1 \circ w_0.
\]
We note that if we def\/ine\footnote{This def\/inition is trivially dif\/ferent to that of \cite{KajiwaraqP3I}, this is to make a full correspondence between the connection preserving deformation above and the translational components of the group of B\"acklund transformations.} $qc^2 = f_0f_1f_2$ and $b_0b_1b_2 = \sqrt{q}$, then the subgroup, $\langle s_0, s_1,s_2, \sigma_0 \rangle$ preserves $c$, while $T_3$ maps $c \to \sqrt{q}c$.

The task remains to make a correspondence between the root lattice and the lattice of connection preserving deformations. However, given the theory established by previous authors~\cite{KajiwaraqP3I}, $T_0$, $T_1$ and $T_2$ present isomorphic evolutions, hence, we should be able to determine six correspondences between the connection  preserving deformations and the lattices. We seek to recover these isomorphisms from the connection preserving deformation standpoint.

Let us list explicitly give the representation of $T_0$:
\[
T_0 : \  \left\{ \begin{array}{c c} b_0 &  b_1 \\ b_2 & \end{array} f_0, f_1, f_2 \right\} \to \left\{ \begin{array}{c c} \sqrt{q}b_0 &  \frac{b_1}{\sqrt{q}} \\ b_2 & \end{array} \tilde{f}_0, \tilde{f}_1, \tilde{f}_2 \right\},
\]
where
\begin{gather*}
\tilde{f}_0f_0  = \frac{qc^2 \left(b_1+f_1\right)}{f_1 \left(b_1 f_1+1\right)},\qquad
\tilde{f}_1f_1  = \frac{qc^2 \left(b_0 \tilde{f}_0+\sqrt{q}\right)}{ \tilde{f}_0 \left(b_0+\sqrt{q}f_0\right)},
\end{gather*}
and $\tilde{f}_2$ is def\/ined by $\tilde{f}_0\tilde{f}_1 \tilde{f}_2 = q c^2$. With regards to $T_0$, we can immediately make a correspondence with $T_{a_1,\lambda_1}^{-1}$ as they are of similar forms. By letting
\begin{gather}\label{T0}
y = - \dfrac{a_1}{b_0f_0},\qquad z= -\dfrac{f_1}{b_1} ,\qquad  b_0^2 = \dfrac{a_1}{a_2},\qquad
b_1^2 = - \dfrac{a_2\kappa_2}{\lambda_1}, \qquad c^2 = \dfrac{a_1a_2\kappa_1}{\sqrt{q}\lambda_1},
\end{gather}
we obtain a correspondence between the evolution in $y$ and $z$ and $f_0$ and $f_1$. The correspondence sought between $T_{\kappa_1,\lambda_1}$ and $T_0$ may be specif\/ied by
\begin{gather}\label{T1}
y =-\dfrac{a_1 b_0 b_1 f_0 f_1}{c^2 q^{3/2}},\qquad z= -\dfrac{f_0}{b_0} ,\qquad  b_0^2 = -\dfrac{a_2\kappa_2}{\lambda_1},\qquad
b_1^2 = -\dfrac{q\lambda_1}{a_1\kappa_2}, \qquad c^2 =\dfrac{a_1a_2\kappa_1}{\sqrt{q}\lambda_1},\!\!
\end{gather}
while the correspondence between $T_{a_2,\lambda_1}^{-1}$ and $T_0$ is specif\/ied by
\begin{gather}\label{T2}
y = -\dfrac{a_2b_1}{f_1} ,\qquad z= \dfrac{q a_1 \kappa _1}{b_0 b_1 f_0 f_1 \kappa _2} ,\qquad  b_0^2 = -\dfrac{q\lambda_1}{a_1\kappa_2},\qquad
b_1^2 = \dfrac{a_1}{a_2}, \qquad c^2 =\dfrac{a_1a_2\kappa_1}{\sqrt{q}\lambda_1}.
\end{gather}
Lastly, there are secondary correspondences between the given lattices. Although \eqref{T0} gives one way of mapping the evolution of $T_{a_1,\lambda_1}^{-1}$ to $T_0$, we also have that
\begin{gather}\label{T0alt}
y = -a_2b_0f_0 ,\qquad z= -\dfrac{1}{b_1f_1} ,\qquad  b_0^2 = \dfrac{a_1}{a_2},\qquad
b_1^2 = -\dfrac{a_2\kappa_2}{\lambda_1}, \qquad c^2 =\dfrac{\lambda_1}{q\sqrt{q}a_1a_2\kappa_1}
\end{gather}
gives another correspondence that inverts the value of $c$. We have similar correspondences that invert $c$ for $T_{a_2,\lambda_1}^{-1}$ and $T_{\kappa_1,\lambda_1}$. The correspondences between the evolutions is summarized in Table~\ref{Correspondences}.

 \begin{table}[!ht]\centering\caption{We list the way in which the lattice of connection preserving deformations may align with the root lattice. The f\/irst three f\/ix $c$ while the last three invert $c$.}\label{Correspondences}

 \vspace{1mm}

\begin{tabular}{| c | c | c | c |}
\hline $T_0$ & $T_1$ & $T_2$ & $T_3$ \\ \hline \hline
\tsep{1pt}\bsep{2pt} $T_{a_1,\lambda_1}^{-1}$ & $T_{\kappa_1,\lambda_1}$ & $T_{a_2,\lambda_1}^{-1}$ & $T_{\kappa_1,\lambda_2}$ \\ \hline
\tsep{1pt}\bsep{2pt} $T_{\kappa_1,\lambda_1}$ & $T_{a_2,\lambda_1}^{-1}$ & $T_{a_1,\lambda_1}^{-1}$ & $T_{\kappa_1,\lambda_2}$  \\ \hline
\tsep{1pt}\bsep{2pt} $T_{a_2,\lambda_1}^{-1}$ & $T_{a_1,\lambda_1}^{-1}$ & $T_{\kappa_1,\lambda_1}$ & $T_{\kappa_1,\lambda_2}$  \\ \hline \hline
\tsep{1pt}\bsep{2pt} $T_{a_1,\lambda_1}^{-1}$ & $T_{\kappa_1,\lambda_1}$ & $T_{a_2,\lambda_1}^{-1}$ & $T_{\kappa_1,\lambda_2}^{-1}$ \\ \hline
\tsep{1pt}\bsep{2pt} $T_{\kappa_1,\lambda_1}$ & $T_{a_2,\lambda_1}^{-1}$ & $T_{a_1,\lambda_1}^{-1}$ & $T_{\kappa_1,\lambda_2}^{-1}$  \\ \hline
\tsep{1pt}\bsep{2pt} $T_{a_2,\lambda_1}^{-1}$ & $T_{a_1,\lambda_1}^{-1}$ & $T_{\kappa_1,\lambda_1}$ & $T_{\kappa_1,\lambda_2}^{-1}$  \\ \hline
\end{tabular}

\end{table}

If we consider the element that changes correspondences between \eqref{T0} and \eqref{T1}:
\[
\tilde{b}_0^2 = \dfrac{a_1}{a_2} = \frac{q}{b_0^2b_1^2} = b_2^2, \qquad \tilde{b}_1^2 = -\dfrac{a_2\kappa_2}{\lambda_1} = b_0^2 , \qquad \tilde{c}^2 = c^2,
\]
similarly
\[
\tilde{f}_0 = \dfrac{a_1}{\tilde{b}_0 y} = \dfrac{q c^2}{f_0f_1} = f_2 , \qquad \tilde{f}_1 = -\tilde{b}_1z = -b_0z = f_0,
\]
hence, the action $b_i \to \tilde{b}_i$ and $f_i \to \tilde{f}_i$ is $\sigma_0$. Similarly, if we consider the element that changes correspondences between \eqref{T0} and \eqref{T0alt}, we have that $\tilde{b}_i = b_i$ and $\tilde{c} = 1/qc$, or more precisely, $q\tilde{c}^2 = 1/qc^2$. The transformations of the $f_i$ are given by
\[
\tilde{f}_0 = \dfrac{a_1}{\tilde{b}_0 y} = \frac{1}{f_0} , \hspace{1cm} \tilde{f}_1 = -\tilde{b}_1z = \frac{1}{f_1}.
\]
The transformation, $b_i \to \tilde{b}_i$ and $f_i \to \tilde{f}_i$ is $\sigma_1$. This adds additional structure to the lattices found in previous studies \cite{Ormerodlattice, OrmerodqPV}.

An interesting secondary consequence is that the above table also presents us with a way in which we may reduce the dimension of the lattice. If there is to be a full correspondence between the connection preserving deformations and the root lattice of type $(A_2+A_1)^{(1)}$ then the composition of the connection preserving deformations that represent $T_0 \circ T_1 \circ T_2$ would be a trivial transformation. Indeed we f\/ind that
\[
T_{a_1,\lambda_1}^{-1} \circ T_{\kappa_1,\lambda_1} \circ T_{a_2,\lambda_1}^{-1} : \ \left\{ \begin{array}{c c c} \kappa_1 & \kappa_2 & a_1 \\ \lambda_1 & \lambda_2 & a_2 \end{array} : w,y,z \right\} \to \left\{ \begin{array}{c c c} \kappa_1 & \kappa_2 & a_1 \\ \lambda_1 & \lambda_2 & a_2 \end{array} : \frac{\kappa_2w}{\kappa_1},\dfrac{y}{q},z \right\},
\]
which corresponds to the identity element in each of the cases presented in Table~\ref{Correspondences}. Hence, we may write
\[
\langle T_0, T_1, T_2 , T_3 \rangle \cong \mathscr{L} / \langle T_{\kappa_1,\kappa_2,\lambda_1,\lambda_2}, T_{a_1,\lambda_1}^{-1} \circ T_{\kappa_1,\lambda_1} \circ T_{a_2,\lambda_1}^{-1} \rangle.
\]
This quotient could be removed by appropriately f\/ixing or removing two redundant variables, which may allow a more explicit correspondence between the variables associated with the linear problem and the $f_i$'s and $b_i$'s.

Using \eqref{T0}, we f\/ind that the symmetry, $S_{a_1,a_2}$ is equivalent to the symmetry
\[
s_1 : \ \left\{ \begin{array}{c c} b_0 &  b_1 \\ b_2 & \end{array} f_0, f_1, f_2 \right\} \to  \left\{ \begin{array}{c c} b_0b_1 &  \frac{1}{b_1} \\ b_2b_1 & \end{array} f_0\frac{b_1 f_1+1}{b_1+f_1}, \tilde{f}_1 \right\},
\]
while using \eqref{T1} we f\/ind $s_0$ and by using \eqref{T2} we f\/ind $s_2$.

\section{Conclusion}

We have that there is additional structure in the lattice of connection preserving deformations as various directions on the lattice present the same evolution equations. By considering the fact that several of the translations on the lattice of connection preserving deformations present equivalent evolution equations, we recover an automorphism group that permutes various directions on the lattice, which corresponds to the group of Dynkin diagram automorphisms. These automorphisms could be incorporated into all the lattices found in the previous study~\cite{Ormerodlattice}. Using these automorphisms, and the one known symmetry, we also recover the symmetries, gi\-ving the full set of B\"acklund transformations from the standpoint of the connection preserving deformations.

\pdfbookmark[1]{References}{ref}
\LastPageEnding

\end{document}